\documentclass[twocolumn,showpacs,preprintnumbers,aps,amsmath,amssymb]{revtex4}

\def\b{\mathbf}

\usepackage{graphicx}% Include Fig.~files
\usepackage{dcolumn}% Align table columns on decimal point
\usepackage{bm}% bold math
\usepackage{color}
\usepackage{epsfig,graphics,amssymb,amsmath,subeqnarray,setspace,graphicx,amsthm,epstopdf,subfigure,txfonts}

\begin{document}

\title{High energy deformation of filaments with internal structure and localized torque-induced melting of DNA}

\author{Arthur A. Evans$^1$}
\author{Alex J. Levine$^{1,2}$}
 
\affiliation{$^1$ Department of Chemistry \& Biochemistry, University of California Los Angeles CA 90095.}
\affiliation{$^2$ Department of Physics \& Astronomy, University of California Los Angeles, CA 90095.}

\date{\today}

\begin{abstract}
We develop a continuum elastic approach to examining the bending mechanics of semiflexible filaments with a local internal degree of freedom that couples to the bending 
modulus. We apply this model to study the nonlinear mechanics of a double stranded DNA oligomer (shorter than its thermal persistence length) whose free ends are linked by 
a single standed DNA chain. This construct, studied by Qu et al. [Europhys. Lett., $\bf{94}$, 18003, 2011], displays nonlinear strain softening associated with the local melting of the double stranded DNA 
under applied torque and serves as a model system with which to study the nonlinear elasticity of DNA under large energy deformations. We show that one can account quantitatively
for the observed bending mechanics using an augmented worm-like chain model, the helix coil worm-like chain. We also predict that the highly bent and partially molten dsDNA should 
exhibit particularly large end-to-end fluctuations associated with the fluctuation of the length of the molten region, and propose appropriate experimental tests.  We suggest that 
the augmented worm-like chain model discussed here is a useful analytic approach to the nonlinear mechanics of DNA or other biopolymer systems.
\end{abstract}


\maketitle

%%%%%%%%%%%%%%%%%%%%%
\section{Introduction}

The worm-like chain (WLC)~\cite{DoiEdwards} has become the standard model of the mechanical behavior of a variety of nanoscale filamentous 
structures in thermal equilibrium.  This approach relies on a single elastic parameter, the bending modulus of the filament, and posits that the bending 
response to an applied torque is simply linear. The well-known and dramatic force vs.\ extension nonlinearities~\cite{MarkoSiggia} of the WLC are fundamentally 
geometric in origin, originating from the inextensibility of the filament.  The bending  mechanics of carbon nanotubes, microtubules, DNA, and other filamentous materials 
can all be modeled as simple elastic rods as long as the local bending deformations are sufficiently small and thus their behavior in equilibrium can be well described by the WLC. 

More recently, it has become clear that, for sufficiently high energy deformations, there are significant discrepancies between the WLC predictions and the 
observed mechanics of such filaments.  For example, carbon nanotubes (CNTs) do not behave as linearly elastic rods at large stresses and strains. When bent sufficiently 
sharply, the competition between bending and stretching in multi-walled CNTs induce rippling modes of 
deformation~\cite{PoplitealWrinkling,Arroyo2008,Arroyo2003,JackmanMeasurement,PoncharalScience1999,PureBendingCNT,bowerCNT} as shown in Fig.~1b. 
The resulting kinks that form in sharply bent CNTs  are reminiscent of the buckled shape that occurs in highly bent drinking straws or tape-measures, a phenomenon 
known as Brazier buckling~\cite{CohenKinks,Brazier1927}. 

Similarly, there is the now significant evidence that biopolymer filaments display more complex elastic behavior that cannot be understood in terms of a simple WLC model.  This 
includes the work on DNA structural transitions induced by applied tension~\cite{CluzelCaron1996,LegerPRL99,StrickBensimon2000,ClausenSchaumannGaub2000_1,ClausenSchaumannGaub2000_2,WennerBloomfield2002,Rouzina2001_1,Rouzina2001_2}, which have been modeled as a type of 
two-state, or helix coil transition~\cite{Ahsan98}. More germane to the current issue are the studies of bending DNA that can be explained only by assuming that the bending 
modulus of DNA is anomalously low in higher energy bends, i.e., ones having a sufficiently small 
radius of curvature~\cite{JieMarkoPRE2003,JieMarkoPRL2004,JieMarkoPRE2005,CriticalTorque,DNAWigginsNano,CompleteDNA,NickedDNA,ProteinChimeras}. One reasonable interpretation of these data 
that is consistent with the mechanical studies of bent carbon nanotubes and macroscopic filamentous  structures is that at high applied stresses the system accesses other 
internal modes of deformation (e.g. rippling of CNT), which, by increasing the number of accessible degrees of 
freedom, necessarily increases the compliance of the filament towards bending. This leads to a nonlinear, stress-induced softening of the material. 

For dsDNA in particular, we recognize there are many such internal degrees of freedom associated with twist, roll, and slide motions of base pairs~\cite{ElHassanCalladine97}, but 
we expect that the principal effect leading to a dramatic reduction in the local bending modulus of the double stranded helical structure is the breaking of 
hydrogen bonds associated with the base pairing interaction and the consequent ``flip-out'' of these individual bases. In essence, we  expect high curvature regions of the chain to 
locally melt into effectively two weakly coupled single stranded chains, which are significantly more compliant to bending. 

It may be possible to make progress on developing more detailed theories of DNA mechanics at short length scales and in high elastic energy configurations by including atomistic
detail in order to capture these effects. This approach, however, sacrifices the elegance and great utility of the WLC model, which accounts for the statistical mechanics of a 
variety of semiflexible filaments using a minimal set of material parameters -- a single bending modulus. We propose that there is a useful intermediate level of description, 
which serves as a minimal extension of the WLC that allows for the observed elastic nonlinearities. 

Our approach is to treat the filament as a two-state elastica~\cite{JieMarkoPRL2004,JieMarkoPRE2005,HCWLC2005,KinkableWigginsPhillipsNelson,NelsonDNAKink,Palmeri2007,Palmeri2008}
in which the filament has a higher bending modulus in the low energy state as compared to the higher energy one. For example the melted, more compliant regions of double-stranded DNA (dsDNA) have
a higher free energy per unit length than the ordered dsDNA, but have a significantly lower bending modulus. This approach to the elasticity of DNA, featuring two local internal states coupled to the conformation of the chain, has been explored previously in studies of the B-DNA to S-DNA transition under tension \cite{CluzelCaron1996,Ahsan98} and in the study of DNA cyclization, where the same local DNA melting as we propose here was implicated in the enhancement of cyclization probabilities \cite{JieMarkoPRL2004}. A similar description of the kinked region of a CNT applies. This assumption 
leads generically to the nonlinear softening of the filament at sufficiently high strains due to the formation of locally compliant regions. In essence, the system exchanges a large 
elastic strain energy per unit length throughout its length to produce a localized internally disrupted region. The energy cost of creating that localized disruption is then 
compensated by the collapse of strain energy into this more compliant part of the filament. 

\begin{figure*}
\includegraphics[width=.75\textwidth]{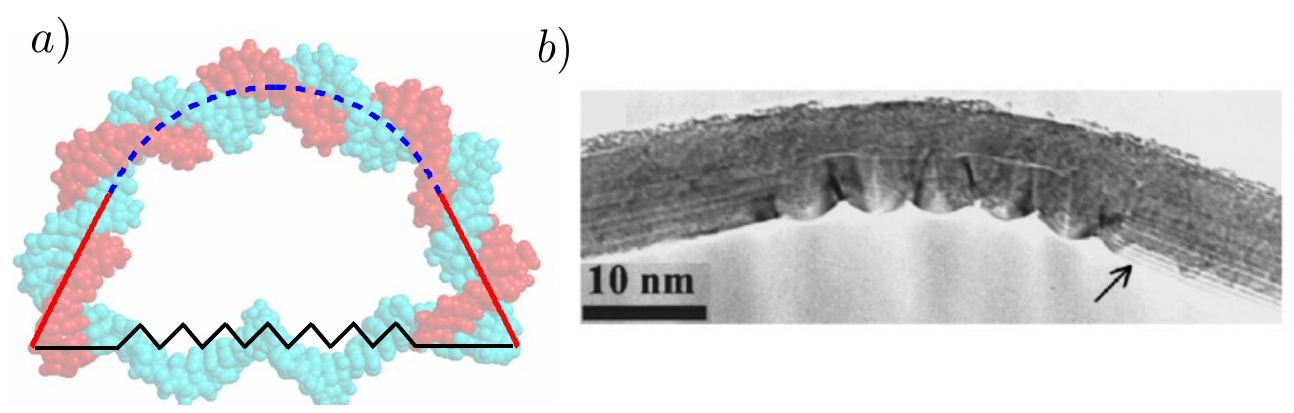}
\caption{(Color online) Internal structure in nano filaments can lead to non-trivial mechanical behavior. a) A hybrid loop of dsDNA and ssDNA can be schematically represented as a filament composed of a stiff (red, solid) region and a softer (blue, dashed) melted region, constrained at it's ends with a spring. b) Under external bending, CNTs develop rippling or kinking modes of deformation that can also be described by postulating phase coexistence between two regions (reproduced from \cite{bowerCNT} with permission). The rippled region has an effectively lower bending rigidity, but there is an energy cost associated with inducing the ripples. }
\label{schematic}
\end{figure*}

In this article, we examine this minimal extension of the WLC model -- the helix/coil wormlike chain (HCWLC)~\cite{HCWLC2005}. This extension requires us to introduce four material parameters. These are the
bending moduli of the filament in its two states, the (free) energy cost per unit length to transform the filament from less compliant to more compliant, and a ``domain wall'' energy
associated with extra (free) energy cost of a boundary between the two internal configuration states of the filament. These latter two parameters are well known in the extensive 
literature on the helix coil model reviewed in Ref.~\cite{Poland}. We then apply this model to examine quantitatively the recent results of Qu et al.~\cite{CompleteDNA} in which they 
constructed a bent piece of dsDNA much shorter than its persistence length by attaching the free ends of this oligomer to single-stranded DNA (ssDNA). In essence, this construct is akin to a bow as used in 
archery, where the tension in the bowstring (ssDNA) is balanced by the bending of the bow (dsDNA); see fig.~\ref{schematic}a. By measuring precisely the elastic energy stored in the bent dsDNA, Qu et al.
obtained a nonlinear torque vs.\ angle curve for the dsDNA oligomer. We use the HCWLC model to interpret these results. It is important to note that the dsDNA in the Qu construct has a nick, a point where the covalently bonded DNA backbone is broken.  This defect serves to localize the generation of the compliant region to the center of the construct, a feature that we exploit when developing our model.

The remainder of this article is organized as follows.  We first develop the general theory in \S II, and then examine the results of applying a constant force to such a 
composite filament (\S III).  We then apply the theory to the experiments of  Qu et al. \cite{CompleteDNA}. We show that one can account for the entire measured elastic 
energy of their construct using well-known values for the thermal persistence length of single and double stranded DNA by adjusting the two remaining helix coil 
parameters in the model. We find the required fitting parameters are consistent with values obtained from other experiments suggesting that the HCWLC model provides a consistent
minimal description of the nonlinear mechanics of DNA, and presumably other filaments having internal structural transitions that can be treated as a two-state internal variable. 
Finally we summarize our results and suggest future directions (\S IV).

\section{Model}

\begin{figure}
\includegraphics[width=.25\textwidth]{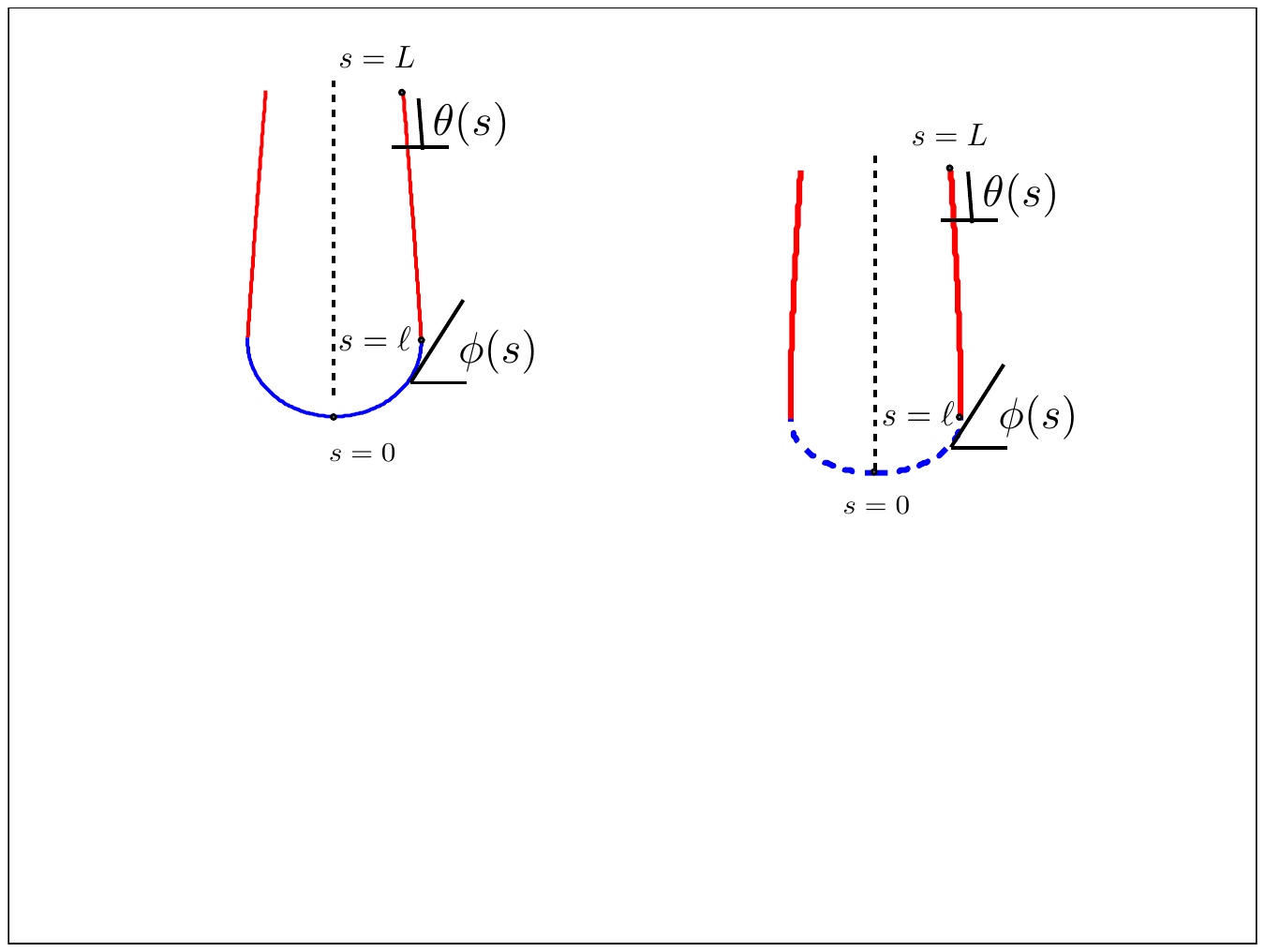}
\caption{(Color online) Schematic of the melting elastica.}
\label{schematicKink}
\end{figure}

In order to capture the essential characteristics of filaments with internal degrees of freedom we use a multi-domain Euler elastica with a prescribed energetic cost for melting (see definition of terms in fig. \ref{schematicKink}).  Our assumption is that, under the influence of external forces or torques, a ``bubble" (or soft region of some variety) may nucleate at the center of the filament, causing a reduction in bending strain at the cost of introducing the melted region.  We treat the two-domain filament as lying in the plane; this will be the lowest energy configuration for the Qu construct, which cannot support twist energy due to the attachment of the ssDNA. Although there should be out of plane fluctuations that contribute to the entropy of the polymer, these will not effect the bending of the stiff region, and thus is irrelevant to the melting transition under study. In 2D we can parameterize the deformation using two tangent angles: $\phi(s)$ and $\theta(s)$ for the soft and stiff regions, respectively.  We let the soft (stiff) region have a bending rigidity denoted by $\kappa_<$($\kappa_>$), and thus the bending energy of the system is given by

\begin{gather}
E_{bend}=\frac{1}{2}\int_0^\ell{\kappa_< \left(\frac{d\phi}{ds}\right)^2}+\frac{1}{2}\int_\ell^L{\kappa_> \left(\frac{d\theta}{ds}\right)^2},
\end{gather}
where $2L$ is the total length of the filament and $2\ell$ is the length of the melted region; we assume here and throughout this work that the system is symmetrical about its midpoint. We augment this energy functional with the internal free energetic cost for melting:

\begin{gather}
E_{int}=H\ell+J,
\end{gather}
where the parameter H is the excess free energy per unit length to be in the thermodynamically disfavored state. For dsDNA (at zero imposed stress), this is the cost of denaturing the filament. For a CNT or other tubelike structure is the cost of introducing ripples or wrinkles. The parameter J is the domain wall energy cost of introducing a boundary between the two internal states of the filament. Because we allow for a continuous melted fraction $\ell$, but only one domain, the energetic cost of J simply shifts calculated criteria for melting.

Without an external stimulus the rod is trivially straight with no melted region, but under the action of an external source non-trivial behavior will appear. By imposing an external compression force F the total energy functional is modified such that

\begin{gather}
E_{tot}=E_{bend}+E_{int}-F\left(\int_0^\ell{\cos\phi ds}+\int_\ell^L{\cos\theta ds}\right),
\end{gather}
leading to equations of equilibrium given by

\begin{gather}
\kappa_<\frac{d^2\phi}{ds^2}+F\sin\phi=0\\
\kappa_>\frac{d^2\theta}{ds^2}+F\sin\theta=0.
\end{gather}
These are the Euler-Lagrange equations for the classical problem of an elastic filament subjected to a constant longitudinal force $F$. 

The appropriate boundary conditions are found by demanding that the boundary terms associated with the variational derivative vanish:

\begin{gather}
\kappa_<\frac{d\phi}{ds}(\ell)=\kappa_>\frac{d\theta}{ds}(\ell)\\
\frac{1}{2}\kappa_<\frac{d\phi}{ds}^2(\ell)-F\cos\phi (\ell)-h=\frac{1}{2}\kappa_>\frac{d\theta}{ds}^2(\ell)-F\cos\theta(\ell)\\
\phi(0)=0\\
\frac{d\theta}{ds}(L)=0
\end{gather}
Physically the first two conditions correspond to the continuity of torque through the interface between soft and stiff regions, as well as continuity of energy density. From a more formal standpoint, these conditions enforce continuity of the canonical momentum and Hamiltonian, and are sometimes referred to as the Weierstrass-Erdmann corner conditions \cite{MomentumBalance}. 
The third boundary condition comes from the assumed mirror symmetry of the system and that there can be only one region of melting, located on this axis of symmetry.  The fourth boundary condition is from our assumption that the force compresses the filament without imposing a torque on the end. General formulations of complicated filaments, including shape-memory alloys \cite{Purohit2002}, adhesive fibers \cite{Carmel2007}, and botanical branches \cite{OReilly2011,FarukSenan2008} often include situations where the continuity of the the conjugate momenta and the Hamiltonian are augmented by a jump condition of some variety; in addition to this the tangent angle need not be continuous through the interface in the presence of shear stresses over the cross-section. For our present purposes it is sufficient to consider that such stresses are contained phenomenologically through the domain wall cost $J$, and we impose $\phi(\ell)=\theta(\ell)$ to complete our formulation. 

By nondimensionalizing lengths by $L$, and rearranging so that the equations of equilibrium are expressed solely in terms of non-dimensional parameters we get the following

\begin{gather}
\frac{d^2\phi}{ds^2}+\Delta^2f\sin\phi=0\\
\nonumber\frac{d^2\theta}{ds^2}+f\sin\theta=0,
\end{gather}
where $\Delta=\sqrt{\kappa_>/\kappa_<}$ is a measure of how drastically the two regions differ in rigidity, and $f=F/F_c$, with $F_c=\kappa_>/L^2$ proportional to the critical force required for classical Euler buckling. Similarly, the boundary conditions become, in terms of these variables,

\begin{gather}
\frac{d\phi}{ds}(\epsilon)=\Delta^2\frac{d\theta}{ds}(\epsilon)\\
\frac{1}{2}\frac{d\phi}{ds}^2(\epsilon)-h=\frac{1}{2}\Delta^2\frac{d\theta}{ds}^2(\epsilon)\\
\phi(0)=0\\
\frac{d\theta}{ds}(1)=0,
\end{gather}
where $\epsilon=\ell/L$ and we have defined $h=HL^2/\kappa_>$ as a dimensionless measure of the energy penalty for introducing the melted region. The reduced energy cost for introducing a domain wall is $j=J/F_cL$. Now, given a set of $\{h,j,\Delta,f\}$ we obtain the size of the melted region that minimizes the energy, denoted by $\epsilon^*$.

\section{Results}

The material parameter space for this particular model is formally fairly expansive, but with an eye towards the DNA problem, and armed with the intuition gained from crumpling and rippling transitions, we can narrow our search to certain regimes. We concentrate on the parameter region $\Delta>1$, since only if there is a significant bending energy gain upon change from melted to unmelted will the nonlinear elasticity regime be accessed. For DNA we estimate the persistence lengths to be $\sim 50 nm$ and $\sim 0.7 nm$ for dsDNA or ssDNA, respectively; as a result $\Delta\sim8$. We consider a broad range of $h$ and $j$, although the energetic cost for denaturing dsDNA has been reported as $\approx 1-1.5 k_BT/bp$, while the domain wall cost is $J\sim 1-10 k_BT$; these two values correspond to dimensionless parameters of $h\sim 3$ and $j\sim 0.1$ \cite{Poland,Ahsan98,HCWLC2005}. For dsDNA we attribute $h$ mainly to the breaking of hydrogen bonds in localized base-pair melting while in CNTs this energy scale corresponds to the work required to produce the short wavelength rippling as shown in fig.~\ref{schematic}b. 

We present the solutions to two problems. First we consider a constant force problem in which a fixed compressive load is attached to the ends. In the second we turn to the experimental system of Qu et al. To do so we include the mechanics of a flexible ssDNA chain connected to the two ends of the much stiffer dsDNA filament. We then compute the elastic energy of this construct as a function of the degree of polymerization of the ssDNA and compare our results to the measurements of Qu et al.

%There are a number of parameters of the model that will specify the particular type of physical system that we wish to examine; the regimes in parameter space that we study will correspond to very different physical phenomena.  Since we are interested in filaments with melted or rippled domains much softer than the exterior region, we look exclusively at $\Delta\gg 1$; this regime leads to the majority of curvature, and thus strain energy, being focused in the melted region.  this behavior is similar to that experienced by crumpled sheets, and exactly the physical behavior we expect from a material that develops a kink.  As a direct result the crucial competition between forces will then be the dimensionless ratio h. For very stiff filaments, with characteristic strains possibly on the order of the thickness, we could expect $h\ll 1$; this is the case of CNTs bent or loaded at their ends.  The ripples that develop in the middle of the tube cost strain energy, but effectively lower the bending resistance \cite{PoplitealWrinkling}.  On the other hand, for filaments that undergo very large deformations without kinking we can expect $h>1$, which is the corresponding limit for dsDNA.  In this case H does not represent the cost of introducing surface undulations, but rather the cost to break the covalent bonds of dsDNA, denaturing the filament and reducing the effective rigidity.

\subsection{Constant force ensemble}

\begin{figure}
\includegraphics[width=.45\textwidth]{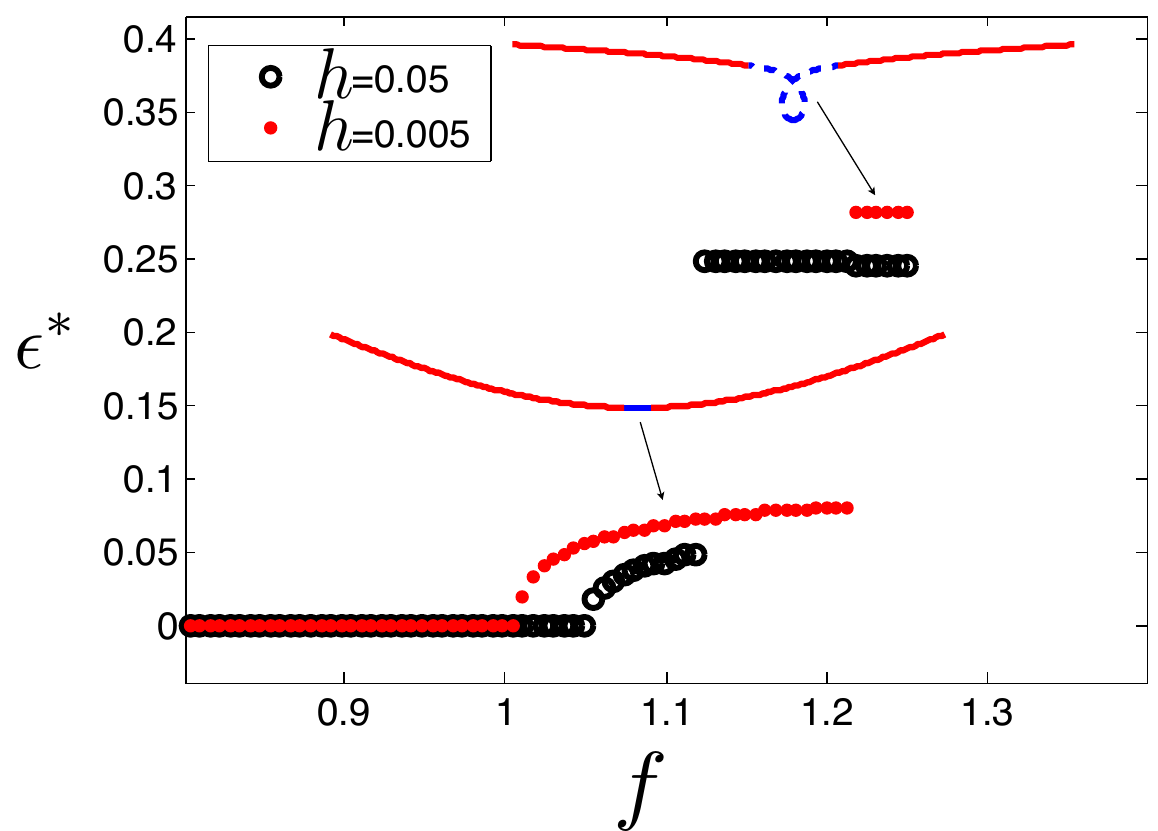}
\caption{(Color online) Melted fraction for the minimum energy configuration as a function of applied force. For small loads the filament does not melt, but at a critical point (dependent on the melting cost $h$) the optimal value for the melted fraction increases continuously from zero.  At yet higher values of the force there is a discontinuous jump in the stable equilibrium point; these solutions, while interesting, are of limited value for our present study and their presence has been included merely for completeness. Parameter values used are $\Delta=8$ and $j=0.15$. }
\label{shapes}
\end{figure}

The simplest result, and perhaps the most intuitively accessible, stems from examining a constant force ensemble. Under a constant load, as is generally the case in experiments on CNTs, we determine the equilibrium conformations resulting from the melting elastica model. One particularly attractive reason for examining this formulation is the ease of computation, as the equations of equilibrium can be solved directly by using Jacobi elliptic functions (see Appendix).

Increasing the applied load (see fig.~\ref{shapes}), we observe a discontinuous melting transition under stress in which the fraction of the chain in the melted state jumps from zero to some finite value at a critical value of the force $f_c=g(h,j,\Delta)$. The dominant balance between the energy required for melting, $h\epsilon$ and the work done by the applied load, $\sim f\cos\theta$ yields the scaling of the critical force to be $f_c\sim h$. At even higher values of the applied force new solutions appear in which the filament makes one or more loops (see fig.~\ref{shapes}), although these solutions are unlikely to be experimentally observable.

We now turn to an analysis of the experimental construct of Qu et al. in which the force applied to the dsDNA is provided by attaching a flexible ssDNA chain to the ends. We compute the experimentally observable elastic energy of the system as a function of the arc length of the ssDNA, i.e. the degree of polymerization.

\subsection{Gaussian chain coupled to the HCWLC}

We treat the ssDNA, which is many times longer than its thermal persistence length $\ell_s$, as a Gaussian chain. Including the lowest order nonlinear corrections we write the elastic energy contribution as \cite{MarkoSiggia}:

\begin{gather}
E_s=\frac{9k_BT}{4N_s\ell_s^2}\left[L_\|^2+\frac{L_\|^3}{N_s\ell_s}+\frac{3L_\|^4}{(N_s\ell_s)^2}\right]
\end{gather}
where 
\begin{gather}
L_\|=\int_0^\ell{\cos\phi}ds+\int_\ell^L{\cos\theta}ds
\end{gather}
is the projected length of the filament in the horizontal direction, $N_s$ is the number of base pairs in the ssDNA. In the limit of small stretching we can modify the energy function given above by the addition of a spring potential:

\begin{gather}
E_{tot}=E_{bend}+E_{int}+\frac{1}{2}k_{ss}L_\|^2,
\end{gather}

 $k_{ss}=\frac{9k_BT}{2N_s\ell_s^2}$ is the spring constant associated with the harmonic part of the stretching energy of the single DNA strand.  For larger values of $L_\|$, approaching non-negligible fractions of the total ssDNA length, including the anharmonic contributions of the Marko-Siggia stretching potential would be necessary.  However, we verify \textit{a posteriori} that we can reproduce the nonlinear elastic effects of Qu et al. using the harmonic contribution alone. We discuss further the role of elastic nonlinearities of the ssDNA chain when considering end-to-end length fluctuations of the construct in section \S III~C.

Now, instead of a constant force ensemble we must deal with the integral constraint on the ends of the filament. Because the stretching energy of the ssDNA enters the energy functional as an integral equation, it is simpler to minimize the energy by constructing the energy landscape numerically and minimizing over the parameter space of choice. We nondimensionalize the energy by $fL$, such that $E/F_cL\sim1$ corresponds roughly to 20$k_BT$ for the relevant values of $\kappa_>$ and $L$. Ideally we wish to compare to the experiments of Qu et al., and thus will examine the elastic energy as a function of $N_s$ (i.e. the shorter the strand, the stiffer the spring). In this scenario, we expect that the energy cost $H\ell$ will correspond roughly to the cost for melting a ``bubble" of DNA, and thus should be on the order of several $k_BT$; similarly, the dimensionless domain wall cost $j=J/fL\sim O(10^{-1})$.  The persistence length of dsDNA is approximately $50$ nm, while that of ssDNA is $\sim 1nm$, leading to a value of $\Delta=\sqrt{\kappa_>/\kappa_<}\sim 8$.

\begin{figure}
\includegraphics[width=.5\textwidth]{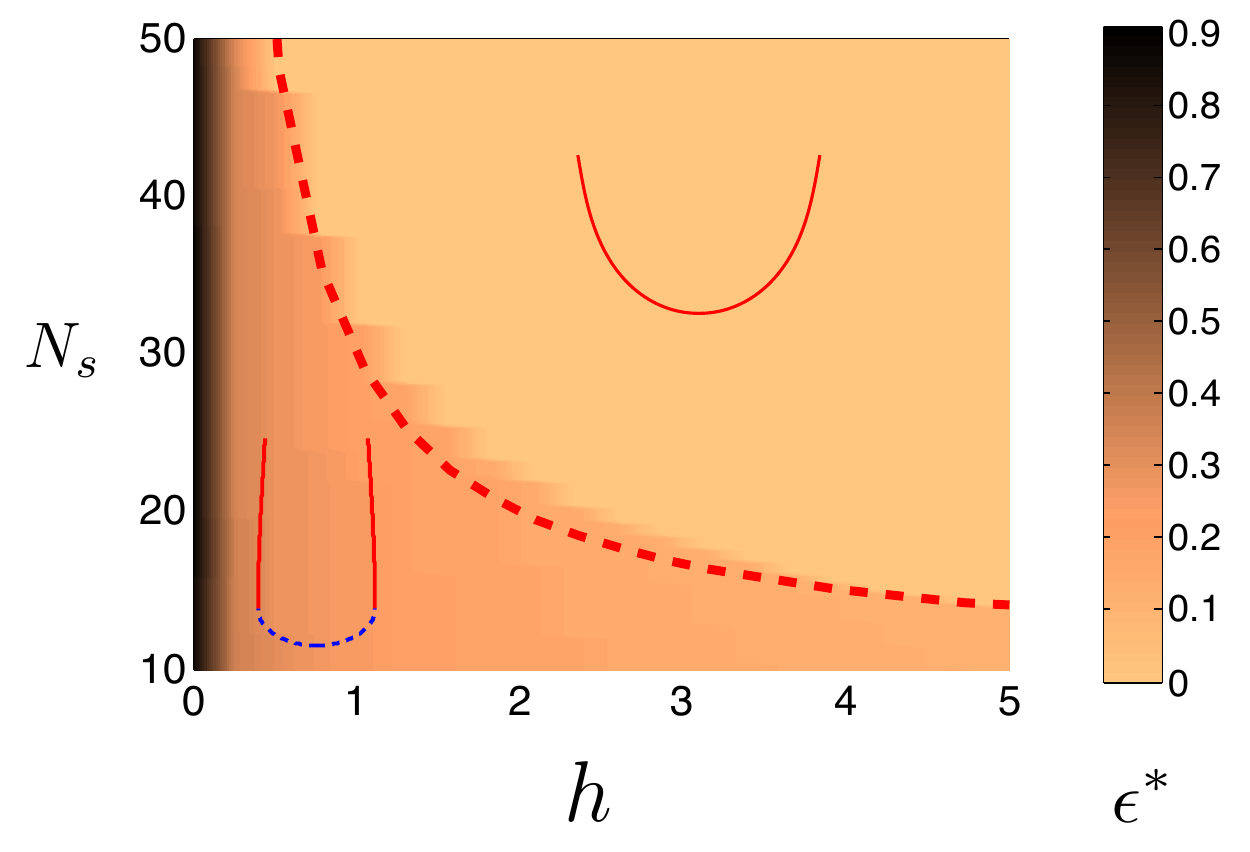}
\caption{(Color online) Phase diagram for melted elastica constrained by a spring. For low values of H the entire filament melts under the action of the spring (black region on left of diagram).  However, for all other values there is a sharp transition between the unmelted and partially melted phase.  Low values of $N_s$ indicate a very stiff spring, and thus the filament can remain partially melted for larger H.  On the other hand, for larger $N_s$ the spring acts only weakly, and we see a very swift transition from completely melted to completely unmelted over a relatively small region of h. The dashed (red) line designates the phase boundary derived from scaling arguments. The parameters used are $\Delta=8$ and $j=0.15$. }
\label{phase_diagram}
\end{figure}

Before discussing the experimental case specifically, we examine more generally the stress induced melting phase diagram for the Qu constructs using a fixed $\Delta=8$, a value appropriate for single- and double-stranded DNA. We also choose a domain wall energy $j\sim 0.15$, well within the range considered appropriate for DNA \cite{Poland,Ahsan98}. The phase diagram, shown in fig.~\ref{phase_diagram}, is spanned by the degree of polymerization of the ssDNA, $N_s$, and the nondimensionalized free energy cost per unit length for being in the more compliant melted phase. The contours reflect equilibrium fraction of the melted dsDNA with darker colors denoting a larger melted fraction. As one might expect, we observe three distinct phases: (i) a completely melted filament obtained for small values of h and $N_s$, (ii) a partially melted or phase coexistence region for intermediate values of the two control parameters, and (iii) for sufficiently high $N_s$ (such that the ssDNA exerts a low compressive load on the dsDNA) or high $h$ (such that the free energy cost for melting is prohibitively high), the chain does not melt at all. We observe a jump in the length of the molten region $\epsilon\sim j^z$ at the phase boundary, and a coexistence boundary that scales as $N_s\sim h^\beta$; here $z$ and $\beta$ are exponents that can be estimated from energy considerations. For the partially melted phase the dominant balance in the energy is between the melting energy and the spring potential, i.e.,

\begin{gather}
h\epsilon+j\sim \frac{1}{N_s}L_\|^2.
\end{gather}
We immediately find the exponents $z=1$ and $\beta=-1$, leading to a phase boundary scaling of $N_s\sim 1/h$, with a position that shifts depending on the value of j.  The dashed (red) line in figure \ref{phase_diagram} shows the scaling behavior is consistent with our calculated phase boundary.

\begin{figure*}
\includegraphics[width=.75\textwidth]{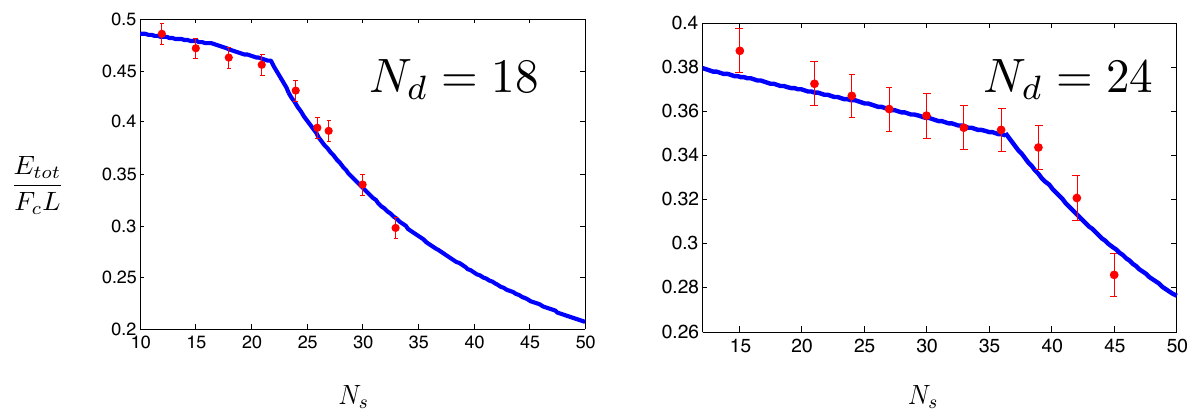}
\caption{(Color online) Total elastic energy of the ssDNA spring and dsDNA system, as a function of the number of ssDNA base pairs $N_s$ with the number of base pairs $N_d=18$ shown in part (a) and $N_d=24$ shown in part (b). The model prediction is shown as the solid (blue) line. The filled (red) points are data from \cite{CompleteDNA}, fit using the parameters $\Delta=8,j=0.15,h=2.1$, for $N_d=18$ and $\Delta=8$, $h=0.78$, $j=0.12$ for $N_d=24$.  In the low $N_s$ region the spring is relatively stiff, leading to a higher melted fraction in the dsDNA and an effective strain softening of the entire construct.}
\label{kink_energy}
\end{figure*}

Our primary result focuses on this sharp transition between the partially melted and unmelted filament; this generic behavior leads to the ability to recreate the nonlinear elastic response measured in tightly bent dsDNA by the experiments of Qu et al.  As seen in figure \ref{kink_energy}, for short springs the energy behaves linearly, while the response for longer springs follows a power law.  Both of these regimes are consistent with the data measured by Qu et al. (red dots in \ref{kink_energy}), and we can use the parameters of the model (the melting cost H and domain cost j) to fit to these data. We find for dsDNA length $\approx 6 nm$ that $h\approx 2.1$, and the value for $j\approx 0.15$. For the case of $N_d=24$, the length of the dsDNA is $K\approx 8nm$ and we find values of $h\approx 0.78$ and $j\approx 0.12$.  Values for $h$ and $j$ reported in the literature range from $h\sim 1.5-3$ and $j\sim 0.05-0.5$~\cite{Poland,Ahsan98,HCWLC2005}. Our model captures the possible sequence dependence of the melting cost $H$, as well as displaying a relatively low value for the longer dsDNA piece, most likely a result of the destabilizing nick at the center of the construct.

\subsection{Effect of fluctuations on melted region}

We expect the phase coexistence region to be characterized by an anomalously soft elasticity. If the end-to-end force on the chain is increased, the chain can accommodate by a combination of bending and increasing the size of the melted region. This soft elasticity associated with the coexistence region should then be reflected in the equilibrium statistics of end-to-end distance fluctuations. In this section, we use our HCWLC model with a melted region to probe these fluctuations and make predictions yet to be tested.

\begin{figure}
\includegraphics[width=.5\textwidth]{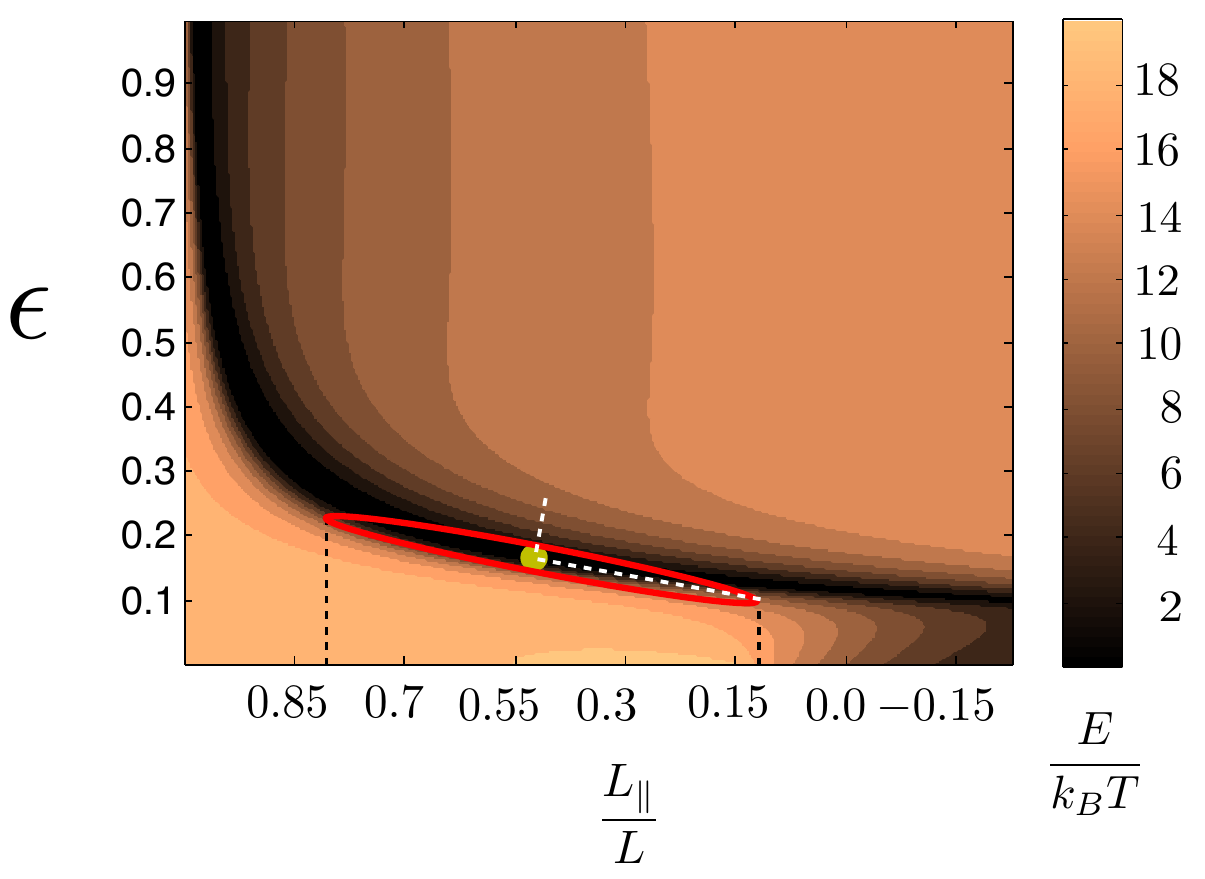}
\caption{(Color online) Energy contours in the space of melted region and projected length.  The (yellow) filled point represents the zero temperature energetic minimum, the (white) dashed lines show the directions of principal curvature, and the (red) elliptical contour corresponds to $k_BT$ energy fluctuations away from the minimum, calculated by expanding the energy landscape to quadratic order about the energy minimum. While the melted region remains relatively constant, the applied force undergoes large fluctuations in the presence of Gaussian thermal excitation.}
\label{fluct}
\end{figure}

To be specific, we compute the end-to-end fluctuations of a Qu construct by examining the Gaussian fluctuations of the end-to-end distance of the dsDNA chain for a fixed polymerization index $N_s$ of the ssDNA, which we continue to treat as a linear spring. Expanding the energy of the system about the point of mechanical equilibrium $(\epsilon^{(0)},L_\|^{(0)})$, we write the energy of the system as

\begin{gather}
\mathcal{H}=\mathcal{H}^{(0)}+\frac{1}{2}\xi_i\xi_j\frac{\partial^2 \mathcal{H^{(0)}}}{\partial\xi_i\partial\xi_j},
\end{gather}
where $\xi_1=L_\|-L_\|^{(0)}$ and $\xi_2=\epsilon-\epsilon^{(0)}$ measure deviations of the size of the melted region and end-to-end distance, respectively. The entire energy landscape of the Qu construct is computed numerically and shown in fig.~\ref{fluct}. The quadratic approximation to the energy is valid near the minimum (shown as the yellow filled circle in fig.~\ref{fluct}) and can be simplified by rotating to the principal axis frame; due to the complex elastic nonlinearities of the system, it is more convenient to compute both the principal axes $\bar{\xi}_1,\bar{\xi}_2$ and corresponding curvatures numerically. In the principal axis frame the Hamiltonian takes the form

\begin{gather}
\mathcal{H}=\mathcal{H}^{(0)}+\frac{\Gamma_1}{2}\bar{\xi}_1^2+\frac{\Gamma_2}{2}\bar{\xi}_2^2,
\end{gather}
where $\Gamma_1,\Gamma_2$ are the principal curvatures associated with the energy landscape. A straightforward application of the equipartition theorem yields the mean squared fluctuations of $\bar{\xi}_1,\bar{\xi}_2$. Using the rotation matrix $U$ relating the principal axis variables to the original axes, $U_{ij}\bar{\b{\xi}}_j=\b{\xi}_i$, we write

\begin{gather}
<(L_\|-L_\|^{(0)})^2>=k_BT\left(\frac{U_{21}^2}{\Gamma_1}+\frac{U_{22}^2}{\Gamma_2}\right).
\end{gather}
Taking experimental parameters ($N_s=15$) and our fitted value of $h=2.1$ ($N_d=18$)we predict that the end-to-end fluctuations of the Qu construct will be quite large.  In fact we find that

\begin{gather}
\frac{\sqrt{<(L_\|-L_\|^{(0)})^2>}}{L_\|^{(0)}}\approx0.6
\end{gather}

For an unmelted dsDNA chain in the Qu construct, the principal curvatures are significantly higher, restricting the end-to-end fluctuations to a much smaller amount: the classical result for the mean square end-to-end distance $<L_\|^2>=L^2f_D(L/\ell_p)$, with $f_D(x)=(1-x+e^{-x})/x^2$ the Debye function \cite{KroyFreyWLC,BauschKroy}. For the dsDNA in the Qu construct $L\ll \ell_p$ and thus the root mean square end-to-end fluctuations $\sqrt{<(L-L_\|)^2>}/L\sim \frac{1}{6}L/\ell_p\approx 0.02$.

One can understand the dramatic softening of the system in the partially melted state as follows.  In the coexistence region, changes in the total bending energy can be partially compensated by adjusting the boundary between the compliant melted region and the stiffer dsDNA. The analogy to the compressibility of a mixture of gas and liquid in the two-phase coexistence region is apt. In both cases the incremental work associated with an increase in the mechanical load remains fixed (instead of increasing), and that work converts material in one phase to the other.

The definitive test of the prediction of the enhancement of the end-to-end distance fluctuations in the Qu construct due to local dsDNA melting is to compare the width of the end-to-end distance distribution for Qu constructs with various ssDNA lengths, $N_s$. Based on the above analysis, we expect one to observe a jump in the width of that distribution as $N_s$ is decreased past $N_s^\star$, where $N_s^\star$ is the critical length of ssDNA below which local melting first occurs.  This length can be independently extracted from the data by finding the break in the slope of the $E_{\rm tot}$ vs. $N_s$ curves, as shown in Fig. \ref{kink_energy}. 

Since the separation of the dsDNA ends in the Qu construct is less than 10nm, FRET pairs attached to the ends of the dsDNA should be accurate reporters of the needed end-to-end distance \cite{StryerHaugland67,Doose2007}, and there are well understood methods available for attaching FRET donor acceptor pairs to DNA. By using single pair fluorescence resonance energy transfer (spFRET) \cite{Deniz1999}, one should be able to extract data on the distribution of this end-to-end distance in freely diffusing Qu constructs in solution.  In fact, this technique has been successfully used to observe the unfolding of a DNA hairpin \cite{Deniz1999}, and thus may be extended to observe the end-to-end distance distribution of the dsDNA in the more complex Qu construct. 

It is important to note that our calculation makes two simplifying assumptions. The first is that the ssDNA can still be treated as a linear spring. It appears that this cannot be valid of the full predicted range of extensions, as the fluctuations in $L_\|$ can become large. The experimental system may in fact be somewhat stiffer for this reason, but the significant fluctuation enhancement of the two-phase regime over that of an unmelted Qu construct remains valid. Nonlinear corrections to the ssDNA energy are not necessary to understand the equilibrium state of the experiments and we do not consider their effect on the energy landscape here, although such corrections are straightforward to pursue numerically. The second assumption used here is that the internal state variable remains in thermal equilibrium throughout the fluctuations.  We expect that this is so, but it may be that the melting/unmelting of the dsDNA will be sequence-dependent and that some sequences may have slower kinetics.

\section{Discussion and summary}

The Qu construct presents an important experimental probe of high energy (small radius of curvature) DNA bending mechanics. In such a regime it is now clear that DNA bending becomes nonlinear
and thus cannot be fully understood in terms of the WLC. This nonlinear softening of dsDNA due to accessing extra local degrees of freedom in high elastic energy deformations, which we associate
with chain melting, is not in itself surprising. Nevertheless, understanding how to account for such effects in an analytically tractable model is essential for the further study of the biophysics of highly 
deformed dsDNA or DNA under large external loads.  To this end, the HCWLC model and it variants have been proposed as natural extensions of the WLC model applicable to such higher energy 
mechanics of DNA and other filaments with internal degrees of freedom, including alpha-helical polypeptides \cite{HCWLC2005,Ahsan98,Palmeri2008,KinkableWigginsPhillipsNelson,NelsonDNAKink}.

We have shown here that a restricted elastica calculation based on the HCWLC is able to account in quantitative detail for the observed nonlinear elasticity of dsDNA as measured by Qu and collaborators. 
To fit the data from the Qu construct using dsDNA with $N_d=18$ or $N_d=24$ base pairs, we fix the thermal persistence lengths of the dsDNA and molten DNA to be $50 nm$ and $\sim 0.7 nm$ consistent with 
measurements of double stranded and single stranded DNA respectively \cite{Tinland,Bouchiat1999}. Using those fixed elastic constants, we find the data of Qu and collaborators can be  
fit by choosing $H\approx .96 k_{\rm B} T/bp$ (for $N_d=18$) or $H\approx 0.2k_BT/bp$, on the same order with that predicted and measured elsewhere \cite{Poland,Ahsan98}.  The effect of the domain wall cost $J$, found to be $J \approx 3k_BT$, is relatively 
independent of our determination of $H$ and compares well to previous helix/coil values of between $1-10 k_BT$ determined from temperature  induced dsDNA melting studies discussed in \cite{Poland,Ahsan98,HCWLC2005}. For the parameters
used, the energy cost associated the bubble dominated by $H$ term , i.e.,  $H \ell \gtrsim J$. For systems in which the compliant domain has an even softer bending modulus, we expect the arc length
of the molten region to be smaller. In this case, the domain wall energy may dominate the length-dependent contribution to the free energy cost of the bubble. 

In addition to showing that the present elastica calculations can account for the dsDNA bending elasticity as measured by Qu and collaborators, we suggest that the presence of a molten bubble leads 
to anomalously large end-to-end distance fluctuations of the Qu construct in the ``two-phase'' regime associated with the experimental range $N_{s} < 25$ -- see Fig. \ref{kink_energy}. These large fluctuations result 
from the correlation between the bending of the dsDNA and change in the arc length of the molten bubble. We propose the observation of a qualitative change in the end-to-end fluctuation 
spectrum of the Qu construct at the kink in the energy versus $N_{s}$ curve represents a key additional test of the theory. The calculations of section IIIC provide a quantitative prediction of that
effect.  In those equilibrium calculations, we have assumed that the internal degrees of freedom of the dsDNA remain in thermal equilibrium on the time scale of the observed end-to-end fluctuations.  While
we expect this is reasonable at present, one can address the potential for nonequilibrium kinetics in the melting and ``refreezing'' of the dsDNA in a manner analogous to that use in the context of 
elastic deformations of shape-memory alloys~\cite{Purohit2002}. 

The principal restrictions made to the full HCWLC model in this calculation were that the applied load produces at most one molten region of the 
dsDNA and that this unstructured part of the molecule is nucleated at the midpoint between the ends of
the dsDNA; it also grows/shrinks symmetrically about that point in response to changes in the applied load. In the case of the experiments in question, 
these assumptions are reasonable. In the case of the former, 
the domain wall energy associated with producing extra molten regions $2 J$ is sufficiently high so as to strongly suppress such fluctuations. To justify the latter, one must recall that the Qu construct
dsDNA is nicked at the midpoint, providing a nucleation center for melting. Moreover, the dsDNA is sufficiently short that the entropic contribution to the free energy associated with the translational 
degree of freedom of the molten bubble is a subdominant correction, as discussed in Ref.~\cite{HCWLC2005}. 

We also do not consider the effect of quenched disorder in this set of calculations. In DNA one expects the local melting temperature to be suppressed in regions where there is a high density of 
AT base pairs. Moreover, there is some evidence that the bending modulus of the molecule also depends on sequence. We have not explored this issue, but one may imagine that the 
position of the kink in Fig. \ref{kink_energy}, corresponding to the nucleation of a molten region and determining our fit for $H$, may depend in detail on the sequence. In addition, the application of this theory to 
microtubules will have to eventually confront the role of quenched disorder since these structures are known to possess a number of defects in local protofilament structure and these are believed
to influence both the zero stress configuration of the filaments and their local elasticity \cite{Fygenson,ChrŽtien1992,Diaz,Nogales,Li2002}. The same can be said for carbon nanotubes \cite{bowerCNT,Arroyo2008,SWDefects,BuongiornoNardelli2000,Falvo1997}. 

This model, although focused on the bending of biopolymers with complex internal structure, can be applied to other types of soft matter elasticity.  Our variational formulation is similar 
in many respects to that employed to describe the peeling apart or adhesion of two or more elastic filaments~\cite{EvansLauga2009,OyharcabalFrisch,Carmel2007} and to discuss 
elastocapillary coalescence~\cite{ElastoCapReview}. As such, it is possible that further application of a multiple domain filament with a measurable cost for peeling could be used in such systems to simplify the analysis.

\section*{Acknowledgments}

We thank G. Zocchi for enlightening conversations, and H. Qu for providing the data used.  Partial support provided by the National Science Foundation (NSF-DMR-1006162).

\appendix

\section{Elastica solutions in terms of elliptic functions}

The elastic energy functional for an elastic rod under uniform compression is

\begin{gather}
E=\int_0^L{\left[\frac{\kappa}{2}\left(\frac{d\theta}{ds}\right)^2-F\cos\theta\right]}ds,
\end{gather}
where L is the length of the rod, $\kappa$ is the flexural rigidity, $\theta$ is an angle that parametrizes the rods tangent vector, $s$ is an arc length coordinate, and $F$ is the compression force. A first variation of this integral yields the Euler-Lagrange equations for the so-called ``elastica":

\begin{gather}
\kappa\frac{d^2\theta}{ds^2}+F\sin\theta=0
\end{gather}
This equation, despite being nonlinear, admits a first integral in $\theta$, indicating that energy is conserved:

\begin{gather}
\frac{1}{2}\left(\frac{d\theta}{ds}\right)^2-F\cos\theta=-F\cos\alpha, \\
\frac{1}{2}\left(\frac{d\theta}{ds}\right)^2-F\cos\theta=-F(1+2\frac{1-k^2}{k^2})
\end{gather}
where we have chosen the constant of integration such that the angle at $s=L$ is identically equal to $\alpha$, for rods with an inflection ($d\theta/ds=0$), and the modulus $k$ is chosen so that filaments without inflections will have a tidy formulation (see below).  There are two main classes of solution, referred to as inflectional (I) and non-inflectional (NI); the former involve solutions that have $d\theta/ds=0$ somewhere on the filament, while the latter do not \cite{Love}.  

Scaling lengths by the filament length L, the solutions for the curvature $d\theta/ds$ everywhere on the filament are given in terms of the elliptic functions $cn(s)$ and $dn(s)$ \cite{AbramowitzStegun}:

\begin{gather}
\frac{d\theta}{ds}= 2k\sqrt{f}\,\, cn\left(\sqrt{f} s+K\right) ,\,\,\,\,\,\,\ (I) \\
\frac{d\theta}{ds}=\frac{2}{k}\sqrt{f} \,\,\  dn\left(\frac{\sqrt{f}s}{k}\right) .\,\,\,\,\,\,\,\,\,\,\,\,\ (NI)
\end{gather}
Here we have defined the reduced force $f=FL^2/\kappa$, the complete elliptic integral of the first kind K, and the Jacobi elliptic functions are defined in the following manner:

\begin{gather}
s=\mathcal{F}(\zeta,k)=\int_0^\zeta{\frac{dt}{\sqrt{1-k^2\sin^2t}}} \\
\zeta=\mathcal{F}^{-1}(s,k) \\
K=\mathcal{F}(\frac{\pi}{2},k)\\
cn(s)=\cos\zeta \\
dn(s)=\sqrt{1-k^2\sin^2\zeta}.
\end{gather}

The angle $\theta$ and the Cartesian coordinates are now given by various integrals,

\begin{gather}
\theta(s)=\int_0^s{\frac{d\theta(s')}{ds'}ds'}\\
x(s)=\int_0^s{\cos\theta(s')}ds'\\
y(s)=\int_0^s{\sin\theta(s')}ds',
\end{gather}
all of which can be expressed in terms of Jacobi elliptic functions or elliptic integrals of various kinds.

Using these general solutions, we need only apply boundary conditions in order to find particular shapes of an elastica under compression. For the case of the melting elastica under a constant compression force we can define the inner melted region as a non-inflectional solution, and then match to the outer filament, which satisfies the inflectional solution.

The boundary conditions are:

\begin{gather}
\phi(L)=0\\
\frac{d\theta}{ds}(0)=0\\
\kappa_>\frac{d\theta}{ds}(\ell)=\kappa_<\frac{d\phi}{ds}(\ell).
\end{gather}
These three conditions come from mirror symmetry at  $s=L$, torquelessness at $s=0$, and torque continuity where the regions change from soft to stiff ($s=\ell$).  For the time being, we assume that the softer region has a fixed length $\ell$, so that our problem reduces to matching two separate solutions for the elastica: $\phi(s)$ satisfies the conditions for (NI) solutions, while $\theta(s)$ corresponds to (I) boundary conditions.

The inflectional solution for $d\theta(s)/ds$ is given by

\begin{gather}
\label{solutions1}
\frac{d\theta}{ds}=2k_1\sqrt{f}\,\, cn(\sqrt{f}s+K(k_1)),  \,\,\,\, 0<s<\epsilon
\end{gather}

Similarly, the non-inflectional solutions for $d\phi(s)/ds$ are given by

\begin{gather}
\label{solutions2}
\frac{d\phi}{ds}=\frac{2}{k_2}\Delta\sqrt{f}\,\,\,\, dn\left(\frac{\Delta\sqrt{f}(1-s)}{k_2}\right), \,\,\,\, \epsilon<s<1
\end{gather}
where $\epsilon=\ell/L$ is the dimensionless position of the domain wall.  Given these two solutions we now need only find the value of the elliptic moduli $k_1$ and $k_2$, which can be found by applying the matching condition (in dimensionless form):

\begin{gather}
\Delta^2\frac{d\theta}{ds}(\epsilon)=\frac{d\phi}{ds}(\epsilon),
\end{gather}

and the condition of force continuity:

\begin{gather}
\Delta^2\left(\frac{d\theta}{ds}(\epsilon)\right)^2=\left(\frac{d\phi}{ds}(\epsilon)\right)^2-2H.
\end{gather}

Using equations \ref{solutions1} and \ref{solutions2} the matching condition can be formulated as a transcendental equation for $k_1$ and $k_2$:

\begin{gather}
\Delta k_1k_2\,\,\,cn\left(f\epsilon+K(k_1^2)\right)=-dn\left(\frac{\Delta f(1-\epsilon) }{k_2}\right).
\end{gather}
The relationship between the two elliptic moduli is given by the second boundary condition, and thus given values for $f$, $\epsilon$, $\Delta$ and $H$ we can now find solutions by solving two transcendental equations. The minimum energy solution can then be found by minimizing the energy with respect to the melted region $\epsilon$.

\bibliography{MeltingElastica}

\end{document}